# Control of surface states of planar metamaterial based on moire effect

**Sergey Yu. Polevoy**[1*] **and Sergey I. Tarapov**[1,2,3**]

[1] *O.Ya. Usikov Institute for Radiophysics and Electronics of NAS of Ukraine,
12 Ac. Proskura St., 61085 Kharkiv, Ukraine*
[2] *Kharkiv National University of Radio Electronics,
14, Nauka Av., 61166 Kharkiv, Ukraine*
[3] *V. Karazin Kharkiv National University,
4, Svoboda Sq., 61022 Kharkiv, Ukraine*

E-mail: [*]polevoy@ire.kharkov.ua, [**]tarapov@ire.kharkov.ua

**Abstract -** The possibility of continuous tuning of the spectral properties of two types of planar metamaterials based on the moire effect by changing their geometric parameters is demonstrated both experimentally and numerically. It is shown that for a one-dimensional moire metamaterial obtained by superposition of two microstrip photonic crystals with close periods, the position of the stop band in the spectrum can be controlled by changing these periods. For the two-dimensional moire metamaterial formed by two identical periodic crossed structures with hexagonal symmetry, the ability to control the frequency of the surface state mode by changing the crossing angle of these structures relative to each other has been experimentally and numerically shown. It is numerically demonstrated that if the moire metamaterial is irradiated by the horn antenna, a surface wave propagating in the metamaterial plane appears in all directions beginning from its intersection point with the axis of the incident wave beam. From the application point of view, moire metamaterials of this type can be considered as promising prototype of microwave filters, whose spectral properties can be continuously and smoothly mechanically rearranged.

## 1. INTRODUCTION

Currently, the interest to studies of the metamaterials based on moire effect occurs. The moire effect consists in the appearance of a pattern when two or more planar structures are superimposed with the same or similar parameters of translational symmetry. In the simplest case we can speak here about the imposition of two identical periodic structures. Thus, the moire metamaterial as a special kind of quasi-periodic structure is formed. Obviously, its electromagnetic properties depend on the crossing angle of structures when they are superimposed (or/and displaced relative each other), on the difference between periods of these structures, etc. This can lead to the appearance of such metamaterial properties that are absent in individual structures that form it. A good illustration of this situation was obtained in the physics of graphene [1], where it was shown that if the crossed graphene layers are superimposed, the certain new energy states are formed on the boundary of the structure.

As well, there is a limited number of papers that study the electromagnetic properties of moire metamaterials in the optical [2,3] and microwave [4] ranges. However, the electromagnetic properties of microwave moire metamaterials are studied not enough stills. This takes place, first of all, because of the complexity of the experiments and difficulties of theoretical description of their topology. In particular, the possibility of implementing continuous/smooth frequency tuning of certain selected surface mode in the metamaterial as a function of the crossing angle of structures forming it is not considered yet.

The aim of this work is to clarify the possibility of control the surface states of some types of planar moire metamaterials by changing their geometric parameters in the microwave frequency range.

## 2. BAND STRUCTURE OF THE SPECTRUM OF ONE-DIMENSIONAL MOIRE METAMATERIAL

### 2.1 Problem statement

As is known, the moire pattern can be formed when two "combs" with close but different period are superimposed. This effect is applied, for example, in the Vernier scale. In order to clarify the principle of control the band structure of the spectrum in one-dimensional moire metamaterials, we chose a structure consisting of two combs in a microstrip design superimposed to each other with a different period. The combs are made from the alternating areas of the microstrip line with different widths that form a spatially limited periodic structure (Fig. 1, a). As it was shown in [5, 6], each of them represents the so-called one-dimensional (1D) photonic crystal (PC) in microstrip design. If the electromagnetic wave excites in such 1D PC with the orientation of the field components indicated on Fig. 1,a, then a set of allowed and stop bands is formed in PCs spectrum [5]. However, when these structures are superimposed, in the case of a small difference in their periods, such a moire metamaterial forms the new kind of quasi-periodic structure (Fig. 1,b).

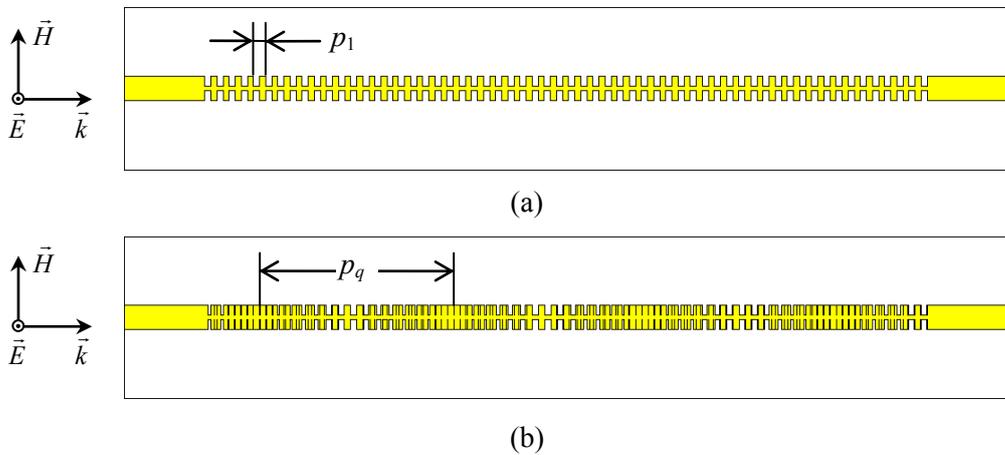

(a)

(b)

**Figure 1.** (a) A schematic representation of the planar structures 90 mm long in a microstrip design: photonic crystal with a period of $p_1$ = 1.5 mm; (b) moire metamaterial formed by superposition of two photonic crystals with periods $p_1$ = 1.5 mm and $p_2$ = 1.6 mm.

It is expected to have slightly different spectrum parameters, which obviously has the band nature. Such moire metamaterial has alternating regions where the average width of the microstrip line is more or less than its average value. Obviously, with a maximum degree of mismatch between wide and narrow sections of the superimposed structures, the average width of the combined structure (the moire metamaterial) is maximal. In the opposite case, i.e. with a minimal degree of mismatch between wide and narrow sections of the superimposed structures, the average width of the moire metamaterial is minimal. It is easy to show that the quasi-period of such moire metamaterial $p_q$ depends on the periods of the forming it PCs $p_1$ and $p_2$, as follows:

$$p_q = \frac{p_1 p_2}{|p_2 - p_1|}. \tag{1}$$

It can be seen from (1), that if the difference in the periods of forming photonic crystals becomes smaller, then the quasi-period becomes larger. Thus, it is possible to choose the periods of subwavelength PCs in such a way that the quasi-period of the obtained moire metamaterial approximately corresponds to one or several half-wavelengths at the selected frequency. In this case, in the spectrum of the transmission coefficient of waves passed through the metamaterial, some new stop bands should be formed. They should lie at significantly lower frequencies than the stop bands of each formative PC.

## 2.2 Calculation results and data analysis

In order to demonstrate the possibility of control the position of the stop band in the spectrum of one-dimensional moire metamaterial, the following parameters were chosen for the numerical calculation: the total length of the structure 90 mm; the period of the first photonic crystal $p_1 = 1.5$ mm; the width of the narrow section of the microstrip line 0.5 mm; the width of the wide section of the microstrip line is 3.0 mm. The lengths of the wide and narrow sections of the microstrip line on each period for both PCs were chosen the same. The metamaterial is placed on the upper side of laminate substrate with a thickness of 1.5 mm. The laminate possesses permittivity is: $\varepsilon' = 3.66$ and $tg\delta = 0.004$ and it is metallized from the lower side. Metal elements are made of copper with a thickness of 0.035 mm. The calculation was carried out by the finite difference method in the time domain (FDTD). Fig. 2 shows the calculated transmission spectrum of such moire metamaterial for several values of period of the second formative PC $p_2$ in the frequency range 0-15 GHz.

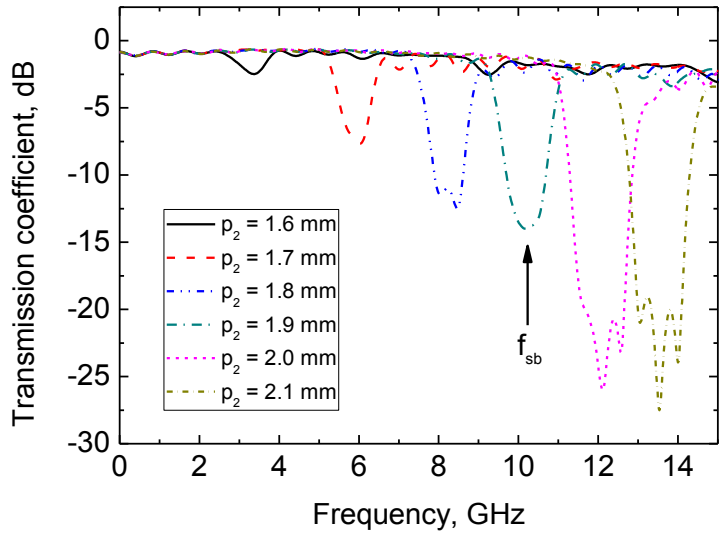

**Figure 2.** Frequency dependence of the transmission coefficient of electromagnetic waves passed through the moire metamaterial in microstrip design, formed by the imposition of two photonic crystals (PC) for several values of the second PC period $p_2$ and with the first PC period $p_1 = 1.5$ mm.

One can see that the dip in the spectrum of the moire metamaterial (the region of the lower value of the transmission coefficient), which corresponds, most likely, to the stop band. It is seen that with the increasing the difference between the periods of the PC, this dip shifts to higher frequencies. To show that this dip really corresponds to the first stop band, the dependence of the middle frequency $f_{sb}$ of the dip on the period of the second PC $p_2$ is presented (Fig. 3, squares). For photonic crystals at the frequency of the first stop band on one period approximately half of wavelength is placed. It is easy to estimate that the frequencies values $f_r$, which correspond to this condition for several values of the moire metamaterial quasi-period $p_q$, are equal to:

$$f_r = \frac{c}{2 p_q \sqrt{\varepsilon_{eff}}}, \qquad (2)$$

where $c$ is the speed of light in vacuum, $\varepsilon_{eff}$ is the effective permittivity of the microstrip line, which for a homogeneous line can be approximately calculated using the formula given in [7]. Applying this formula (for a homogeneous microstrip line) with a strip width $W = 3$ mm, we obtain $\varepsilon_{eff} \approx 2.82$. Let us construct the dependence of the frequency on the period of the second PC $p_2 - f_r(p_2)$ for two values of permittivity: $\varepsilon_{eff} = 2.82$ (Fig. 3, triangles) and for $\varepsilon_{eff} = 4.0$ (Fig. 3, circles). It is easy to see that for the case of $\varepsilon_{eff} = 4.0$ the calculated frequency $f_r$ well approximates the revealed middle frequencies $f_{sb}$ of the band gap. Consequently, the wavelength in the structure at the frequency of the middle of the stop band is smaller than in a homogeneous microstrip line.

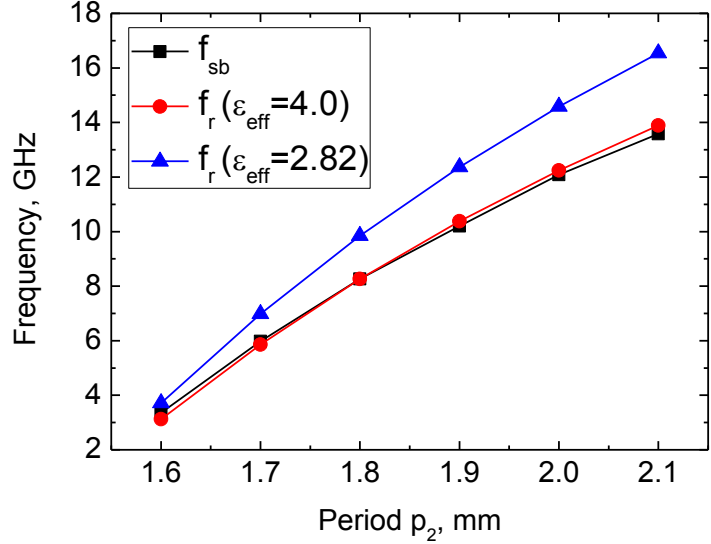

**Figure 3.** The frequency of the middle of the stop band $f_{sb}$ in the transmission spectrum of the moire metamaterial formed by the imposition of two photonic crystals (PC) on the period $p_2$ of the second PC value.

Note that with an increase in the period of the second photonic crystal (Fig. 2), the depth of the stop band gap also increases. This is easily can be explained by the fact that with an increase in the period of the second PC, the quasi-period decreases according to formula (1), and at a fixed length of the resulting moire metamaterial the number of such quasi-periods increases also. And as is well known, the depth of the stop band of a spatially bounded photonic crystal is in the first approximation proportional to this number.

Thus, the frequency position of the stop band in the spectrum of 1D moire metamaterial in microstrip design can be effectively controlled by varying the periods of 1D photonic crystals, which form it.

## 3. SURFACE STATES IN A TWO-DIMENSIONAL MOIRE METAMATERIAL

### 3.1 Problem statement

Another way to design a moire pattern is the superposition of two identical periodic planar structures crossed at some angle. In order to realize the control of frequencies of surface states in two-dimensional metamaterials based on the moire effect, two superimposed at a certain angle structures with hexagonal symmetry have been chosen. Each structure presents itself a two-periodic array of thin metal equilateral hexagons with side $a$, located in the nodes of a hexagonal lattice with a period $p$ (Fig. 4, a). When two such structures are superimposed with some crossing angle $\alpha$, a quasi-periodic structure is formed in the moire metamaterial (Fig. 4, b). Easy to see, that the unit cell of structure also has a hexagonal symmetry (as in the initial structures), but it has a larger size.

In different points of such metamaterial, the different distances between the centers of the metal hexagons of two formative structures observe. With the greatest distances between the centers of metal hexagons of different structures, the area of mutual overlap is the minimal. And at the smallest distances between the centers of metallic hexagons, they almost overlap with each other (Fig. 4, b, the central region). It can be shown that the quasi-period $p_q$ of such combined structure (the hexagonal moire metamaterial) depends by the period of the formative structures $p$, and for a small crossing angle $\alpha$ it is equal [1] to:

$$p_q = \frac{p/2}{\sin(\alpha/2)}. \qquad (3)$$

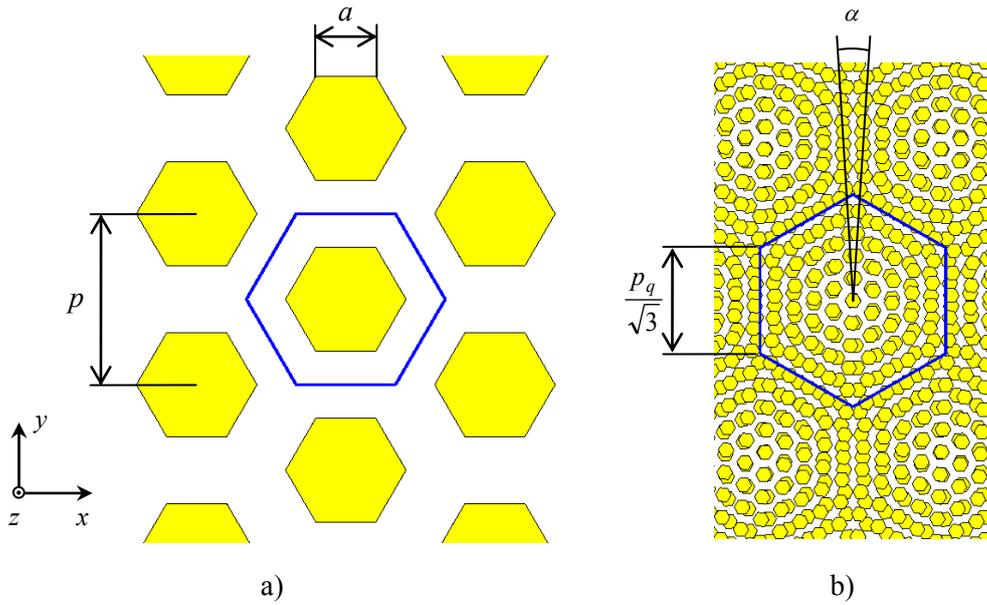

a)            b)

**Figure 4.** (a) A schematic representation of elements of the moire metamaterial formed by the superimposing two identical planar structures with crossing angle $\alpha$: the unit cell of each planar structure; (b) the unit cell of the quasi-periodic structure of the moire metamaterial at a crossing angle of $\alpha = 7°$. The hexagonal unit cells are marked by bold lines (blue - online).

From formula (3) it can be seen that with decrease of the crossing angle of the formative structures, the value of the quasi-period increases. Our numerical calculation has showed that in a certain range of ratios between the side of hexagon $a$ and the period of the forming structure $p$ in the hexagonal moire metamaterial, almost continuous metallic stripes of irregular shape ("current paths") are formed along the unit cells borders of the moire metamaterial (Fig. 4, b). As a result, the surface modes appear in the metamaterial, and this process has the percolative character.

In particular, if the $a/p$ values are too large, the metal stripes are very wide, up to the case that most of the metallic elements overlap with each other. In such a structure there are practically no conditions for the formation of the flow of large closed (ring-like) conductive currents on the surface of the moire metamaterial. Therefore, no surface modes arise here.

Further, as $a/p$ decreases, at a certain threshold value (the "upper percolation threshold"), the above-mentioned metal strips of irregular shape ("current paths") are quite narrow and almost continuous, so closed conductive currents of appreciable magnitude can flow on moire metamaterial surface. Then in the moire metamaterial the surface modes of noticeable magnitude arise.

Subsequently, with a decrease of $a/p$, at certain "lower percolation threshold", there is a tendency to the formation of gaps in metal strips at certain crossing angles of the structures. As a result, the intensity of the surface modes decreases sharply.

Our numerical calculations have showed that for the moire metamaterial of such type there is only a narrow range of $a/p$ values (approximately found as 0.33–0.38) at which surface states (modes) appear. In this range, in the transmission spectrum the gaps, whose frequencies depend on the crossing angle of the formative structures, should occur. We will discuss this phenomenon in more detail in the next subsection.

### 3.2 Numerical evaluation of frequency-angular characteristics of surface states in moire metamaterial

In order to control the frequencies of surface states in two-dimensional moire metamaterials based on superimposed structures, the following parameters were chosen: the period of the hexagonal lattice of formative structures is $p = 2$ mm, the size of the side of copper hexagons is $a = 0.7$ mm, and their thickness is 0.035 mm. Structures are put close to each other with a crossing angle $\alpha$. The calculation was carried out by the finite difference in the time domain (FDTD) for the frequency range of 1–15 GHz and crossing angles range $\alpha = 3\text{-}12°$ (Fig. 5) with a normal incidence of electromagnetic waves.

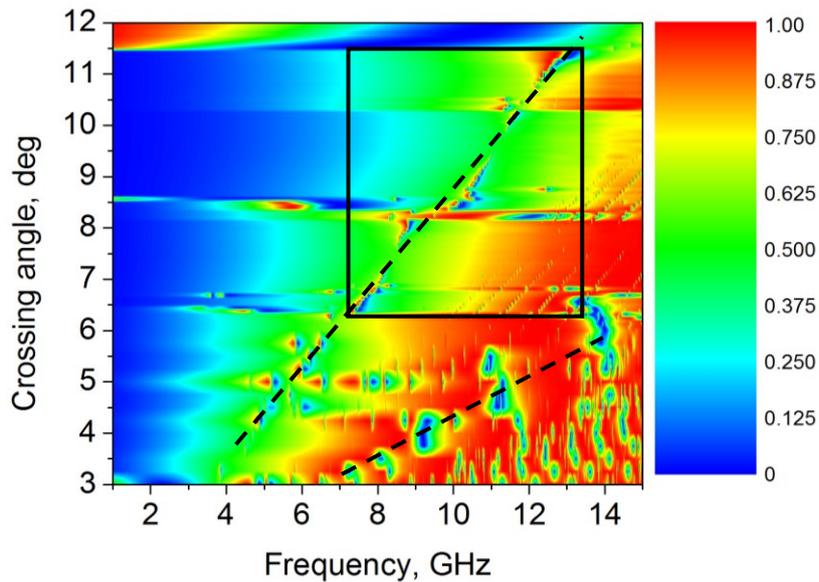

**Figure 5.** The transmission coefficient of electromagnetic waves passed through the moire metamaterial, in dependence of the frequency and crossing angle of the formative structures. Dashed lines mark the most intensive modes of surface states.

One can see from Fig. 5 that in a certain range of frequencies and crossing angles of structures (marked as rectangular area in the figure) there is a resonant minimum for transmission of waves, which shifts to higher frequencies with the crossing angle increasing. This resonance shift should be explained by the fact that the length of the abovementioned formed metal strips ("currents paths") depends on the crossing angle between the structures. This resonance in the spectrum corresponds to one of the surface state modes in this structure. Note that in the range of angles of 8.5-11.5° a monotonous increasing the frequency of this surface state mode is observed with increasing of crossing angle. Note, that other, more high-frequency resonance gaps are also observed in the spectrum. They are caused by the flowing of current through other, shorter paths in the metamaterial. However, they have a significantly lower intensity and a narrower range of tuning angles.

Note that while irradiating such hexagonal moire metamaterials with a spherical wave front (for example, from a horn antenna) a kind of very interesting surface waves at certain resonant frequencies may be formed. They propagate from the point on the metamaterial, which is closest to the irradiation source.

At the same time there is a minimum of the transmission coefficient in the spectrum is observed, and a significant part of the incident energy passes to the surface wave. It is shown in Fig. 6 an example of the distribution of the surface current magnitude for moire metamaterial formed by two hexagonal structures with a period $p = 2$ mm crossed at angle $\alpha = 10°$ at the resonant frequency with normal incidence of waves from the rectangular horn antenna with an aperture of 33 mm × 47 mm from a distance of 100 mm to the metamaterial plane.

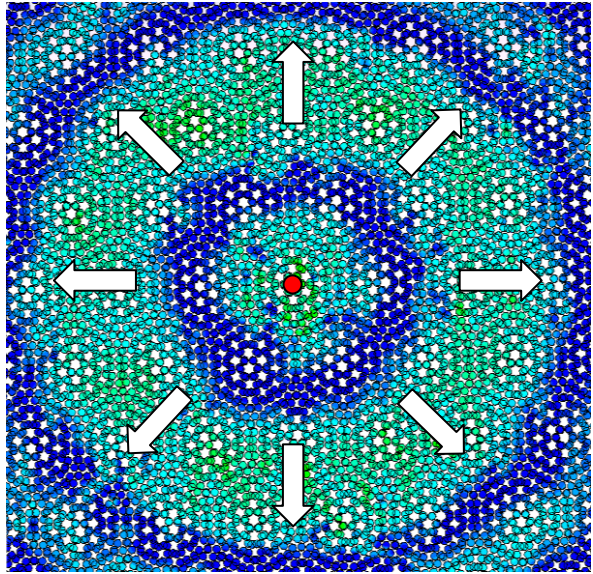

**Figure 6.** The distribution of the surface current amplitude for the moire metamaterial, formed by the superposition of two identical planar hexagonal structures crossed at an angle $\alpha = 10^\circ$, with its normal irradiation by the horn antenna.

Thus, it can be seen that at the resonant frequency at a certain crossing angle of the formative structures in the moire metamaterial, a surface wave is formed. It spreads in concentric circles from the center of the structure (red dot in Fig. 6), through which the axis of the horn antenna passes. The arrows indicate the direction of propagation of the surface wave.

### 3.3 Test experiments and data analysis

With the purpose of experimental verification of the revealed resonant properties of two-dimensional metamaterials used the moire effect, the hexagonal planar structure has been fabricated (Fig. 7, a). The period of the hexagonal lattice of formative structures is chosen as $p = 1.125$ mm and the size of the side of metal hexagons is $a = 0.375$ mm. The hexagons are made of copper with a thickness of 0.035 mm, placed on a layer of FR-4 laminate ($\varepsilon' = 4.35$, $tg\delta = 0.022$) with a thickness of 1.5 mm. The metamaterial consisted of two structures that put close to each other by the sides with copper hexagons at some crossing angles (which can be varied smoothly). It is important to ensure the absence of air gaps between the structures, since they strongly influence the magnitude of the resonance minimum in the spectrum.

The experimental setup (Fig. 7, b) consists of the metamaterial under study with a size of 100 mm ×100 mm placed between two irradiating rectangular horn antennas [8]. Aperture of horns is 33 mm × 47 mm, and their length equals 100 mm. Antennas are located on the same axis that passes perpendicularly to the metamaterial plane through its center. The measurement circuit is fitted to the Vector Network Analyzer *Agilent N5230A* by coaxial-waveguide adaptors and coaxial cables. Using the Network Analyzer, the parameter $S_{21}$, which in fact is the transmission coefficient for waves passed normally through the metamaterial in the frequency range 7...15 GHz, is registered.

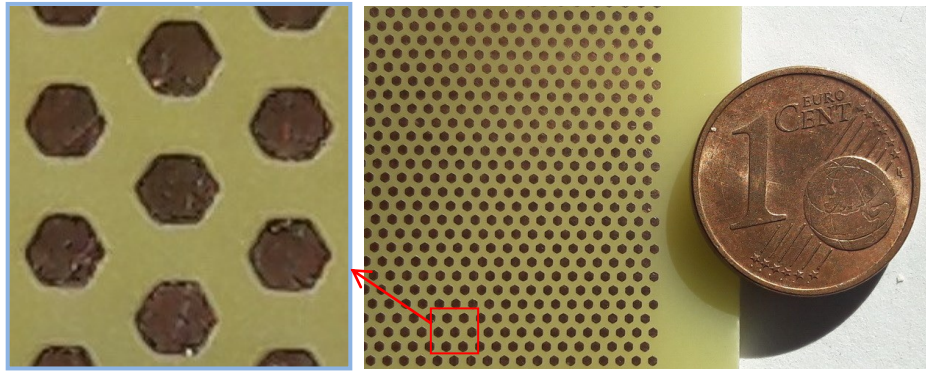

(a)

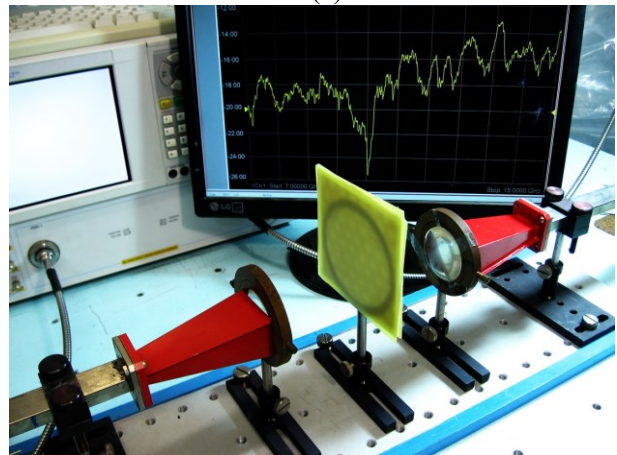

(b)

**Figure 7.** (a) The view of a two-periodic array of equilateral thin metal hexagons on a dielectric (FR-4 laminate) substrate, that used as one of the crossed (superimposed) structures for the moire metamaterial; (b) The view of the experimental setup to study the moire metamaterial.

In order to compare the experimental data with the results of numerical simulation, the frequency-angular dependence of the transmission coefficient of wave with normal incidence passed through the metamaterial was registered (Fig. 8). Red circles mark those theoretical values of frequencies of the minimum transmission coefficient at several selected crossing angles of the formative structures, which were then compared with the experimentally obtained data.

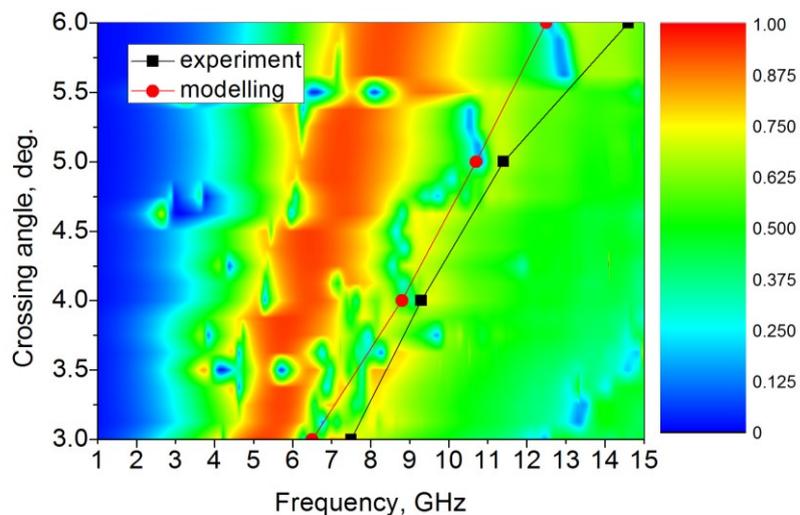

**Figure 8.** Experimental and calculated frequency of the surface state mode in the hexagonal moire metamaterial, formed by the imposition of two identical crossed planar structures, depending on the angle of their crossing.

During measuring the resonant frequencies, which are caused by the appearance of the surface states, some certain control crossing angles of the formative structures were selected. It can be seen that there is good qualitative agreement between the experimentally obtained frequencies of the surface state (Fig. 8, black squares) and the results of numerical simulation (Fig. 8, red circles).

Note that, as follows from the results of numerical simulation, the depth of the resonant minimum can be increased by using a substrate material with lower losses for this frequency range.

## 4. CONCLUSIONS

Thus, the possibility of continuous tuning of the spectral properties of two types of planar metamaterials based on the moire effect by changing their geometric parameters was experimentally and numerically demonstrated. It has been shown that:

1. The quasi-periodic moire metamaterial, obtained by superposition of two one-dimensional periodic photonic crystals in microstrip design with close periods, demonstrates an unexpectedly deep stop bands in the transmission spectrum. In this case, the position of the stop band can be controlled by varying periods of photonic crystals, which forming the moire metamaterial.

2. For the quasi-periodic metamaterial formed by two identical superimposed crossing structures with hexagonal symmetry, for a certain range of ratios of the sizes of its elements, the resonance minima, corresponding to surface states are observed in the transmission spectrum. The possibility to control the frequency of such surface state by varying the crossing angle between the formative structures has been experimentally and numerically shown. The range of ratio between the size and period of elements in the formative structures, at which the frequency tunable surface state in moire metamaterial can be excited, has been numerically determined.

3. It is numerically demonstrated that at certain parameters of studied moire metamaterial, irradiated with horn antenna, the surface wave spreading in the plane of the metamaterial in all directions from the intersection point with the axis of incident wave beam takes place.

In conclusion, we note that from an application point of view, metamaterials of such a type can become the prototype of microwave filters, whose spectral characteristics can continuously and smoothly mechanically tuned. Note, that the implementation of magnetoactive elements in such structures is of special interest.

## REFERENCES

1. Fleischmann, M., Gupta, R., Weckbecker, D., Landgraf, W., Pankratov, O., Meded, V., and Shallcross, S., "Moiré edge states in twisted graphene nanoribbons," *Phys. Rev. B.*, Vol. 97, 20, 205128, 2018.
2. Wu, Z., and Zheng, Y., "Moiré Chiral Metamaterials," *Advanced Optical Materials*, Vol. 5, 16, 1700034, 2017.
3. Wu, Z., Liu, Y., Hilla, E.H., and Zheng, Y., "Chiral metamaterials via Moiré stacking," *Nanoscale*, Vol. 10, 38, pp. 18096-18112, 2018.
4. Han, J.-H., Kim, I., Ryu, J.-W., Kim, J., Cho, J.-H., Yim, G.-S., Park, H.-S., Min, B., and Choi, M., "Rotationally reconfigurable metamaterials based on moiré phenomenon," *Opt. Express*, Vol. 23, 13, 17443-17449, 2015.
5. Belozorov, D.P., Girich, A.A., and Tarapov, S.I., "Analogue of surface Tamm states in periodic structures on the base of microstrip waveguides," *The Radio Science Bulletin*, Vol. 345, 64-72, 2013.
6. Belozorov, D.P., Girich, A.A., Nedukh, S.V., Moskaltsova, A.N. and Tarapov, S.I., "Microwave Analogue of Tamm States in Periodic Chain-Like Structures," *Progress in Electromagnetics Research Letters*, Vol. 46, 7-12, 2014.
7. Pozar, D. M., *Microwave engineering*, 4th ed., John Wiley & Sons, Inc., 2012.
8. Polevoy, S.Yu., "An experimental technique for estimating constitutive parameters of chiral media in the millimeter wavelength range," *Telecommunications and Radio Engineering*, Vol. 73, No. 8, 681-693, 2014.